\documentclass[prl,amsmath,amssymb,aps,twocolumn]{revtex4}
\usepackage[T1]{fontenc}
\usepackage[utf8]{inputenc}
\usepackage{lmodern}
\usepackage{graphicx}
\usepackage[colorlinks=true,linkcolor=blue,urlcolor=blue,citecolor=blue]{hyperref}

\usepackage{color}
\usepackage{xcolor}
\usepackage{times}
\definecolor{dg}{rgb}{0.1, 0.7, 0.1}

\newcommand{\bea}{\begin{eqnarray}}
\newcommand{\eea}{\end{eqnarray}}
\newcommand{\beq}{\begin{equation}}
\newcommand{\eeq}{\end{equation}}
\newcommand{\be}{\begin{equation}}
\newcommand{\ee}{\end{equation}}

\newcommand{\rme}{\mathrm{e}}
\newcommand{\rmd}{\mathrm{d}}
\newcommand{\nn}{\nonumber}
\renewcommand{\epsilon}{\varepsilon}
\newcommand{\nott}[1]{}

\newcommand{\Fig}[1]{\includegraphics[width=\columnwidth]{./#1}}

\begin{document}

\renewcommand{\omega}{\gamma}

\title{%
Distribution of velocities   in an avalanche, and related quantities: Theory and numerical verification} 

\author{Alejandro B. Kolton${}^{1}$, Pierre Le Doussal${}^{2}$, Kay J\"org Wiese${}^{2}$}
\affiliation{${}^{1}$ Centro At\'omico Bariloche and Instituto Balseiro,
Comisi\'on Nacional de Energ\'ia At\'omica (CNEA),
Consejo Nacional de Investigaciones Cient\'ificas y T\'ecnicas (CONICET),
Universidad Nacional de Cuyo (UNCUYO),
Av.\ E.\ Bustillo 9500, R8402AGP San Carlos de Bariloche, R\'io Negro, Argentina\\
\mbox{${}^{2}$Laboratoire de Physique de l'Ecole Normale Sup\'erieure, ENS, Universit\'e PSL,} CNRS, Sorbonne Universit\'e, Universit\'e Paris-Diderot, Sorbonne Paris Cit\'e, 24 rue Lhomond, 75005 Paris, France.} 

\begin{abstract}
{ We study several probability distributions relevant to the avalanche dynamics of elastic interfaces
driven on a random substrate: The distribution of size, duration, lateral extension or area, as well as velocities. 
Results from the functional renormalization group and scaling relations involving
two independent exponents, roughness $\zeta$, and dynamics $z$, 
are confronted to high-precision numerical simulations of an elastic line with short-range elasticity, i.e. of internal dimension $d=1$.
The latter are based on a  novel stochastic algorithm which generates its disorder on the fly. 
Its precision grows linearly in the time discretization step, and 
it is parallelizable. 
Our 
results show good agreement between theory and numerics, both for the critical exponents as for the scaling functions.
In particular, the   prediction ${\sf a} = 2 - \frac{2}{d+ \zeta - z}$ for the   velocity exponent is confirmed with   good accuracy. }
\end{abstract}
\maketitle

\begin{figure}[bth]
\Fig{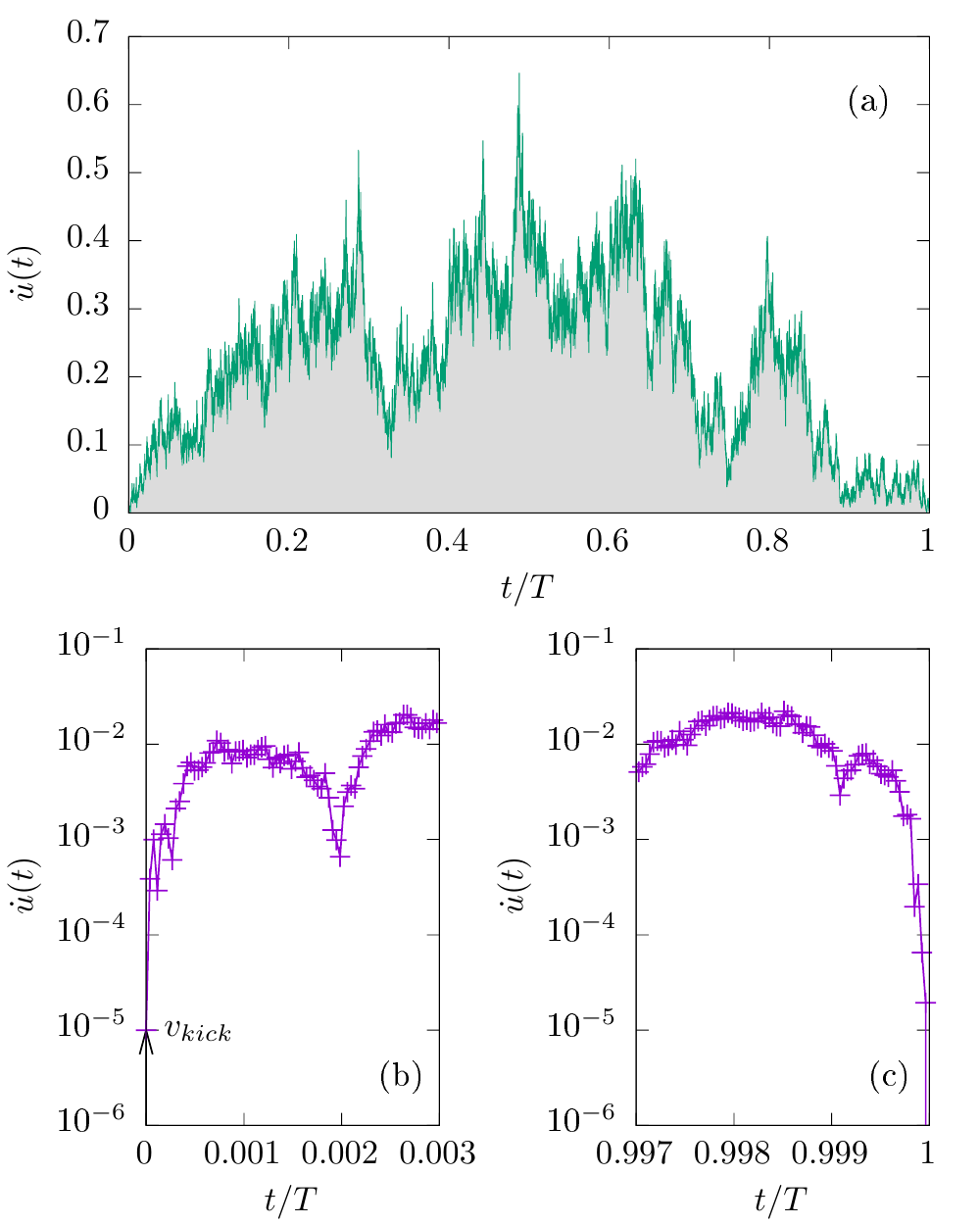}
\caption{(a) The velocity as a function of time $t$ for one   avalanche of duration $T$. (Parameters are $T=26.2$,  $A=10$, $L=64$, $m=1$). Zooms of the departure (b), and ending (c) of the avalanche. The arrow in (b) indicates the magnitude $v_{\rm kick}$ of the uniform velocity kick triggering the avalanche.} \label{fig:samplesconfigs}
\end{figure}

\paragraph{Introduction}
Disordered systems, when driven slowly or via a small kick, do not respond smoothly, but in a bursty way. An example are elastic manifolds, or more specifically   elastic strings,  subject to a random potential. 
 An example for the  global  velocity as function of time is shown  on figure \ref{fig:samplesconfigs}. 
 At $t=0$, the system received a small kick. The velocity as a function of time $t$ then performs a random walk, which terminates at a precise moment in time. Driving the system adiabatically slowly, it is at rest for most of the time, interceded with jerky motion as the one shown on figure \ref{fig:samplesconfigs}. Each such event is called an avalanche. 
  Avalanches are   ubiquitous,  found in   earthquakes in geoscience \cite{GutenbergRichter1956}, Barkhausen noise \cite{Barkhausen1919,CizeauZapperiDurinStanley1997} in dirty disordered magnets, contact-lines on a disordered substrate \cite{LeDoussalWieseMoulinetRolley2009}, and many more.

The theory has been developed for many years, starting with phenomenological and mean-field arguments \cite{BuldyrevBarabasiCasertaHavlinStanleyVicsek1992,DSFisher1998}. In the context of magnets, a more systematic approach was proposed by Alessandro, Beatrice, Bertotti, and Montorsi (ABBM) \cite{AlessandroBeatriceBertottiMontorsi1990b,AlessandroBeatriceBertottiMontorsi1990}, who reduced the equation of motion for a magnetic domain wall to a {\em single degree of freedom}, subject to a random force modeled as a random walk. It was only  realized later  \cite{LeDoussalWiese2011a} that the Brownian force model (BFM) is the correct mean-field theory for avalanche dynamics. In contrast to the ABBM  or similar mean-field models, which have a single degree of freedom,  the BFM is an extended  model, in which each degree of freedom, i.e.\ each piece of the elastic manifold, sees an independent random force, which itself is a random walk. In \cite{LeDoussalWiese2011a} it was shown that its center of mass is the same stochastic process as the single degree of freedom of the ABBM model.

The BFM is  the  starting point for a  field theory of elastic manifolds subject to short-ranged disorder. It allows to calculate a plethora of observables beyond pure scaling exponents, as e.g.\ the full size \cite{LeDoussalWiese2011b,LeDoussalWiese2009a,LeDoussalWiese2008c} distribution, the  velocity distribution \cite{LeDoussalWiese2012a,LeDoussalWiese2011a}, and the temporal \cite{DobrinevskiLeDoussalWiese2011b,DobrinevskiLeDoussalWieseTBP} or spatial shape \cite{ThieryLeDoussalWiese2015,DelormeLeDoussalWiese2016,AragonKoltonDoussalWieseJagla2016,ZhuWiese2017} of an avalanche.

In this letter we study numerically, and compare to field theory, the distributions of the duration $T$ of an avalanche, its size $S=\int_{0}^{T}\dot u(t) \rmd t$, its velocity   $ \dot u $, and extension $\ell$.
We  briefly review the corresponding scaling relations, and then confront them to large-scale numerical simulations.  
The latter has been possible through the development of a novel powerful algorithm 
which generates its disorder on the fly 
by accurately solving an extension of the BFM to incorporate short-ranged disorder. 
\begin{table*}[t]
\newcommand{\n}{\hspace*{0.15mm}}
\scalebox{1.0}{\begin{tabular}{|c|c|c|c|c|}
\hline
 \n&\n $P(S) $ \n&\n   $P(T)$ \n&\n   $P(\dot u)$ \n&\n   $P(V)$ \\
\hline
\hline 
    \n&\n $S^{-\tau}$ \n&\n  $T^{-\alpha}$  \n&\n  $\dot u^{-\sf a}$      \n&\n $  V^{-{\sf k}_{V}} \rule[-1ex]{0mm}{3.5ex}$ \\
\hline 
 SR elasticity     \n&\n $\tau = 2 - \frac2{d+\zeta}$   \n&\n
 $\alpha = 1+ \frac{d-2+\zeta}{z}$ \n&\n   ${\sf a} = 2- \frac2{d+\zeta-z}$ \n&\n   $ {\sf k}_{V} =  2- \frac{2-\zeta}d  \rule[-1.5ex]{0mm}{4.5ex} $  \\
\hline 
LR elasticity   \n&\n$\tau = 2 - \frac1{d+\zeta}$\n &  \n$\alpha = 1+ \frac{d-1+\zeta}{z}$\n& \n${\sf a} = 2- \frac1{d+\zeta-z}$\n & \n  $ {\sf k}_{V} = 2- \frac{1-\zeta}d \rule[-1.5ex]{0mm}{4.5ex}$ \\
\hline 
\end{tabular}}\hfill
\renewcommand{\n}{\hspace*{0.15mm}}
\scalebox{1}{\begin{tabular}{|c|c||c|c||c|c|c|c|c|}
\hline 
  \n &  \n$d$\n & \n$\zeta$\n & \n$z$\n & \n$\tau$\n &  \n$\alpha$\n &  \n${\sf a}$\n  & \n$\gamma$  \n &\n  ${\sf k}_{V}$\n \\
\hline 
\hline
\n &  \n$1$\n & \n$5/4$\n & \n$10/7$\n & \n$ 10/9$ \n &  \n$47/40$\n &   \n$-10/23$\n & \n$1.57$  \n &\n   1.25\n\\
\cline{2-9}
\n SR\n &  \n$2$\n & \n$0.75$\n & \n$1.56$\n & \n$1.27$\n &   \n$1.48$\n &  \n$0.32$\n  & \n$1.76$  \n &\n   1.38\n\\
\cline{2-9}
\n &  \n$3$\n & \n$0.35$\n & \n$1.75$ \n & \n$1.40$\n &   \n$1.77$\n &  \n$0.75$\n & \n$1.91$  \n &\n  1.45\n\\
\hline
\n LR\n &  \n$1$\n & \n$0.39$\n & \n$0.77$ \n & \n$1.28$\n &   \n$1.51$\n  &   \n$0.39$\n& \n$1.81$  \n &\n   1.39\n \\
\hline
\end{tabular}}
\caption{Left: Scaling relations. 
Right: Critical exponents obtained via the scaling relations using standard values
for $\zeta$ and $z$ \cite{LeschhornNattermannStepanowTang1997,DuemmerKrauth2005,FerreroBustingorryKolton2012,GrassbergerDharMohanty2016}.}
\label{tab1}
\end{table*}

\paragraph{Definition of the model.}
Consider the over-damped equation of motion for a manifold in a random-field environment, 
\beq\label{2}
\partial_{t} u_{x,t} = \partial_x^2  u_{x,t} + F(x,u_{x,t}) + m^2 (w_{t}-u_{x,t})\ .
\eeq
The manifold is trapped in a harmonic potential of width $m^2$, and position $w_{t}$. The well is moved  slowly, either via $w_{t}=v t$ in  the limit $v\to 0^+$ (constant velocity driving), or by augmenting $w$ by a small amount $\delta w$ at discrete times $t$ (kick driving). $F$ is a short-ranged correlated random force, which will be specified below. Eq.~(\ref{2}) 
 provides a well-defined framework to study avalanches, both in field theory \cite{LeDoussalWiese2011a,DobrinevskiLeDoussalWiese2011b,LeDoussalWiese2012a,DobrinevskiLeDoussalWiese2014a,DobrinevskiPhD,DobrinevskiLeDoussalWieseTBP}, as for simulations 
\cite{RossoLeDoussalWiese2006a,
RossoLeDoussalWiese2009a,
KoltonSchehrLeDoussal2009,
FerreroBustingorryKolton2013,
FerreroBustingorryKoltonRosso2013,
KoltonJagla2018,
CaoBouzatKoltonRosso2018}.
Indeed, the velocity as a function of time performs a burst-like evolution, with a well-defined beginning and end, see figure \ref{fig:samplesconfigs}, separated by periods without activity (not shown).  

\begin{figure*}
\Fig{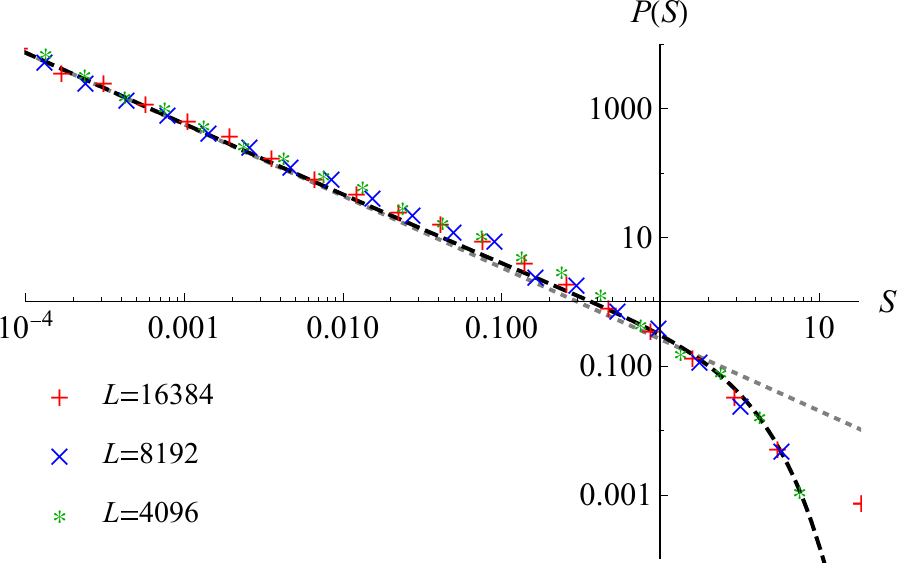}~\Fig{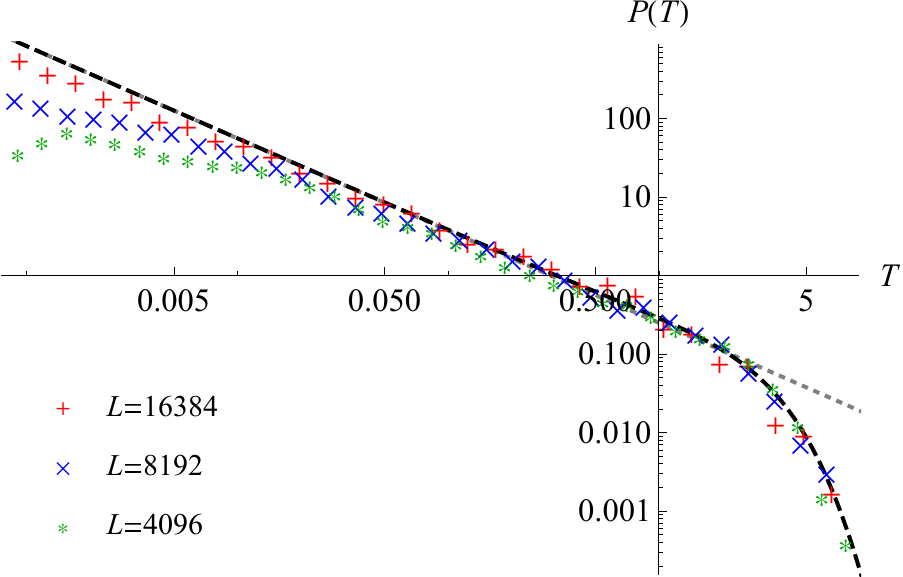}
\caption{Left: The rescaled distribution of size $P(S)$. 
To avoid system-spanning avalanches, the ratio $L m=10$ is kept fixed.  The black dashed line is the 1-loop result (12) of Ref.~\cite{LeDoussalMiddletonWiese2008}, the gray dotted line the pure power law.   
Right: {\em ibid} for the duration distribution $P(T)$.  The analytical result  is given in Eqs.~(3.113)-(3.116) of Ref.~\cite{DobrinevskiPhD} and in Ref. \cite{DobrinevskiLeDoussalWieseTBP}.}\label{f:PofS}
\end{figure*}%
\begin{figure}
\Fig{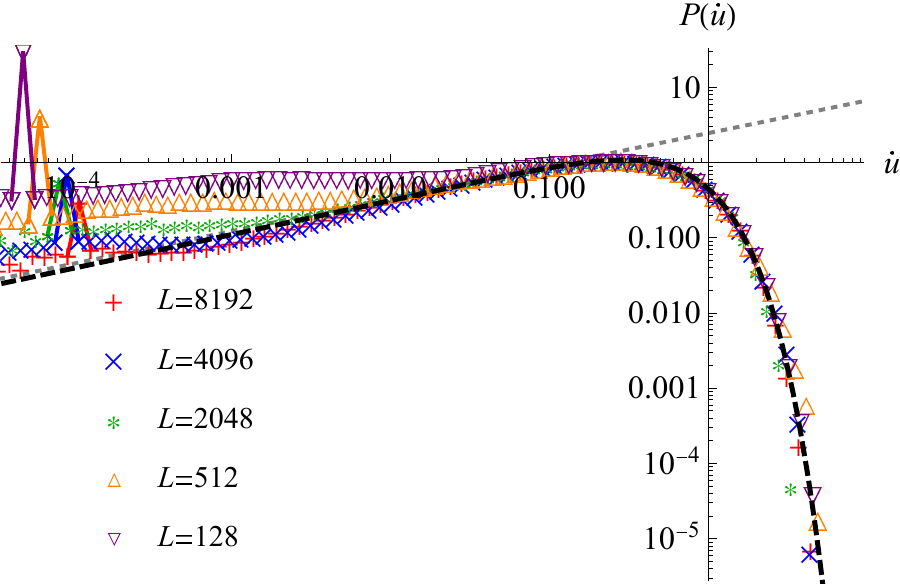}
\caption{The center-of mass velocity distribution $P(\dot u)$. The  weight of the peak at $\dot u=v_{\rm kick}$  is $\frac{\delta t}{\left< T\right>}\sim L^{-z}\sim m^z$, where $T$ is the duration of an avalanche and $\delta t$ the time discretization step.  
The analytic result (black dashed line) is from Eq.~(385) of Ref.~\cite{LeDoussalWiese2012a}, the dotted gray line the pure power law $P(\dot u)\sim u^{{\sf a}}$. }
\label{f:Pofv}
\end{figure}%

\paragraph{Scaling relations.}
The theory of the depinning transition of interfaces \cite{NattermannShort,NarayanDSFisher1992b,NarayanDSFisherInterface,
LeschhornNattermannStepanowTang1997,ChauveLeDoussalWiese2000a,ChauveLeDoussalWieseDepLong}
introduces two independent critical
exponents, the roughness exponent $\zeta$, and the dynamic exponent $z$. Within the field theory developed in 
\cite{LeDoussalWiese2011a,DobrinevskiLeDoussalWiese2011b,LeDoussalWiese2012a,DobrinevskiLeDoussalWiese2014a,DobrinevskiPhD,DobrinevskiLeDoussalWieseTBP} no independent exponent is required to describe avalanches. As a consequence, their  exponents   are given by scaling relations, together with
the requirement of the existence of a massless field theory \cite{DobrinevskiLeDoussalWiese2014a},
a generalization of the arguments of   Ref.~\cite{NarayanDSFisher1992b}. 
Consider the PDF $P_{\delta w}(S)$ of the total
size $S$ of the avalanche following a small kick $\delta w$.
Its large-size cutoff $S_m \sim m^{-(d+\zeta)}$ is defined through the ratio of its first two moments
\cite{LeDoussalWiese2008c}
\be \label{Sm} 
S_m = \frac{\left<S^2\right> }{2\left<S\right>} \ .
\ee 
The PDF reads, for $S$ larger than a microscopic cutoff
\be \label{PDFS} 
P_{\delta w}(S) \simeq \frac{\left<S\right>}{S_m^2} p(S/S_m) \quad , \quad \left<S\right> = \delta w L^d\ ,
\ee
where $p(s)$ is a universal scaling function with $p(s) \sim s^{-\tau}$ at small $s$,
defining the size exponent $\tau$. Existence of a massless field theory imposes that
 the avalanche density per unit applied
force, $\rho_f(S)= \lim_{\delta w \to 0} \frac1{m^2 \delta w} P_{\delta w}(S)$
has a finite limit for $m \to 0$. This requires $m^{-2} S_m^{\tau-2}$  to be independent
of $m$ at small $m$, hence $\tau=2 - \frac{2}{d+\zeta}$, recovering the value of Narayan and Fisher \cite{NarayanDSFisher1992b}. The exponents for
the avalanche duration $T$, or lateral size $\ell$ are then obtained by writing $P_{\delta w}(S) \rmd S  = P_{\delta w}(T) \rmd T$, and using the appropriate scaling relations between $S$, $m$  and $T$,
leading to the results in the Table \ref{tab1}, where numerical values are given as well. 
We also consider the (spatially integrated) velocity at time $t$ after the kick, $\dot u(t)= \int \rmd^d x\, \partial_t u_{x,t}$.
The PDF of the total velocity $\dot u=\dot u(t)$ is obtained by considering many successive
kicks and sampling the time $t$ uniformly. Its associated density is
$\rho_f(\dot u) \sim \int \rmd t\, \rho_{f}\big(\dot u(t)\big)$. By scaling it takes  at small $\dot u$ the form
$\rho_f(\dot u)=\frac{L^d}{m^2 v_m^2} (v_m/\dot u)^{\sf a}$ where ${\sf a}$ is the 
velocity exponent, $v_m=S_m/\tau_m$ and $\tau_m \sim m^{-z}$ is the large time cutoff.
Requirement of a massless limit implies 
\be
{\sf a} = 2 - \frac{2}{d+ \zeta - z}\ ,
\ee 
a main prediction of Ref.\ \cite{DobrinevskiLeDoussalWiese2014a}, in agreement
with the $\epsilon$ expansion of Ref.\ \cite{LeDoussalWiese2012a}, and 
which we test numerically below.

\paragraph{The algorithm: Theory.}
The equation of motion of an elastic manifold due to short-ranged disorder-forces can be generated by the following set of equations (with an arbitrary constant $A$) \cite{DobrinevskiPhD,LeDoussalWiese2014a}  
\bea \label{1}
&& \partial_{t} {\cal F}_{x,t} = -A {\cal F}_{x,t} {\dot u_{x,t}} + \sqrt{2 A  \dot u_{x,t}} \;\xi(x,t) \ , \\
\label{2b}
&& \partial_{t} \dot u_{x,t} = \partial_x^2  \dot u_{x,t} + \partial_{t}{\cal F}_{x,t} + m^2 (v-\dot u_{x,t}) \ ,~~~~~~\\
&& \left< \xi(x,t) \xi(x',t') \right> = \delta (x-t')\delta(t-t')\ .
\eea
Rewriting $F_{x,t}$ as a function of $x$ and $u_{x,t}$, ${\cal F}_{x,t}\equiv F(x,u_{xt})$ yields for each $x$ an evolution equation of $F(x,u)$, 
\begin{eqnarray}
  \partial_u {  F}(x,u ) = -A {  F}(x,u ) + \sqrt{2A} \;\eta(x,u)\ , \\
  \left<  \eta(x,u)  \eta(x',u') \right> = \delta(x-x')\delta(u-u')\ .
\end{eqnarray}
The solution to this equation is 
\begin{eqnarray}
  \overline{{  F}(x,u){  F}(x',u')}  
  &=& 
   \delta(x-x') \,\rme^{-A|u'-u|}\ .
\end{eqnarray}
We can read off the microscopic disorder force-force correlator
\beq
 \Delta(u-u') = \rme^{-A|u'-u|}\ .
\eeq
It is short-ranged, and {\em microscopically rough}. 
The problem is how to solve efficiently the stochastic   equations  (\ref{1})-(\ref{2b}). Discretizing time in steps of size $\delta t$ yields
\bea
 \dot{u}_{t+\delta t} - \dot{u}_{t} &=& 
{\cal F}_{t+\delta t}-{\cal F}_{t}+O(\delta t) \nn\\
& =&  \sqrt {2 A \dot u_{t} \delta t}\, \xi_{t} + O(\delta t)\ ,
\eea
where $\xi_{t}$ is a normal-distributed Gaussian random variable with mean $\left< \xi_{t}\right>=0$, and variance $\left< \xi_{t} \xi_{t'}\right>=\delta_{t,t'}$. 
Since one is interested in the limit of $\delta t\to 0$, the appearance of $\sqrt {  \delta t}$  in front of the noise term implies a rather slow convergence.

\paragraph{The algorithm: An Improved Solver.}
The idea is to solve analytically the  random process 
\beq
\partial_{t} \dot u_{t} = \sqrt{2 A  \dot u_{t}}\, \xi(t) 
\eeq
with absorbing boundary conditions at $\dot u=0$ for a finite interval $\delta t$.  Following  \cite{DornicChateMunoz2005}, we first write the analytic solution of the corresponding Fokker-Planck equation 
\bea
P(\dot u,t) &=& \delta(\dot u) \exp\left(-\frac{\dot u_0}{At}\right) \nn\\
&&+ \frac{\exp\left(-\frac{\dot u_0+\dot u}{At}\right) }{At}\sqrt{\frac{\dot u_0}{\dot u}}\, I_1\!\left(\frac{2 \sqrt{\dot u_0 \dot u}}{A t}\right) \ ,~~~
\eea
where $I_{1}$ is the Bessel-$I$ function of the first kind. 
It can be reexpressed as a series
\beq
P(\dot u,t) = 
\sum_{n=0}^{\infty} p_n \,\frac{1}{A\,t}P_n \!\left(\frac{\dot u}{A t}\right)
\eeq
with 
\bea
p_n &=& \frac{\left(\frac{\dot u_0}{A t}\right)^n \exp\left(-\frac{\dot u_0}{A t}\right)}{n!}\\
P_0(y) &=& \delta(y) \\
P_n(y)&=& \frac{y^{n-1} \exp\left(-y\right)}{(n-1)!}\ , \qquad n\ge 1\ .
\eea
The algorithm   consists  of two steps: First draw a random number $n$ from a Poisson distribution with parameter $\frac{\dot u_0}{A t}$. In a second step draw a random number $y$ from a Gamma-distribution with the (previously determined) parameter $n$.  This   yields 
\beq
\dot u_{t+\delta t} = \dot u_{t} + A\,  y\, \delta t\ , 
\eeq
to which have to be added the drift terms proportional to $\delta t$.

\begin{figure}
\Fig{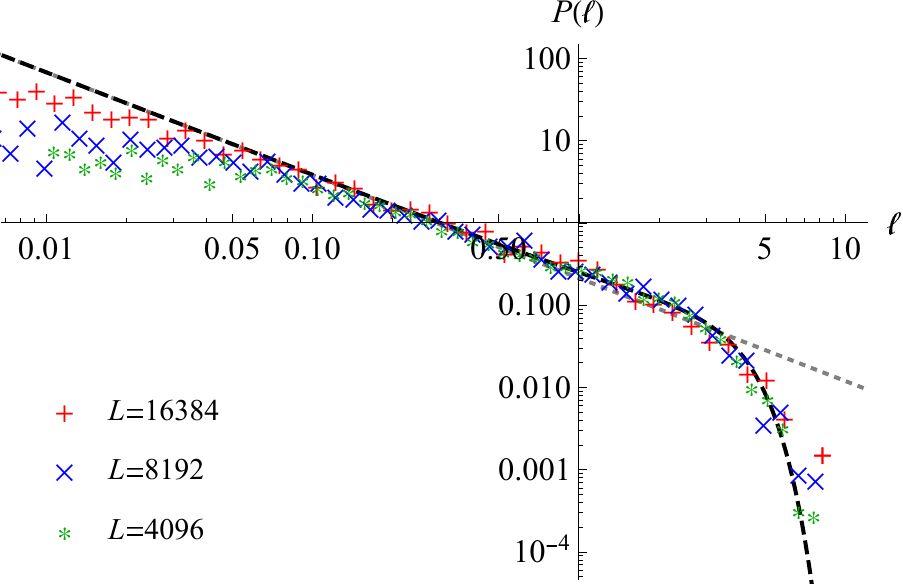}
\caption{The distribution of lateral sizes $P(\ell)$. In absence of analytic results for the scaling function, we use the relation $P(\ell )\rmd \ell = P(S) \rmd S$, and $S=\ell^{{d+\zeta}}$ to infer the latter (black dashed line). A pure power law $P(S)\sim S^{{-\sf k}}$ is given as the gray dotted line. }
\label{f:PofL2}
\end{figure}%
\paragraph{Results: Size and duration distributions.}

Our simulations are performed in dimension $d=1$. 
In figure \ref{f:PofS}, we report our findings for the avalanche-size and duration distributions, both known analytically from the $\epsilon=4-d$ expansion \cite{LeDoussalWiese2008c,DobrinevskiPhD,DobrinevskiLeDoussalWieseTBP}. The size distribution was also checked numerically \cite{RossoLeDoussalWiese2009a}. One extends the
definitions \eqref{Sm} and \eqref{PDFS} to observables $\cal O$ such as the duration $T$ and 
extension $\ell$ (see below) by writing the PDF
\beq\label{scalingform}
P_{\delta w} ({\cal O}) = \frac{\left< {\cal O} \right> }{{\cal O}_m ^2} 
p\!\left( \frac{{\cal O}}{{\cal O}_m }\right)\ ,
\eeq
where ${\cal O}_m = \frac{\left<{\cal O}^2\right> }{2\left<\cal O\right>}$ is
the characteristic large-scale cutoff and 
$p(x)$ is a universal function depending only on $d$ and ${\cal O}$,
such that $\int_0^{\infty} \rmd x\, x p(x)=1$,  
$\int_0^{\infty} \rmd x\, x^2 p(x)=2$. It is this function
$p(x)$ which is plotted in figure \ref{f:PofS} and \ref{f:PofL2}
from our simulation, (denoted there by $P(x)$)
and compared to its prediction from the $\epsilon$ expansion
(via an extrapolation to $\epsilon=3$). 
While the scaling relations using $\zeta=5/4$ and $z=10/7$ predict a size exponent $\tau = 1.11$
and a duration exponent $\alpha = 1.17$,  see table \ref{tab1},
our best fits are 
\bea
\tau &=&  
1.2 \pm  0.2 \ ,\\
\alpha &=&  1.1 \pm 0.15\ .
\eea
\paragraph{Velocity distribution.}
For the center of mass, the velocity distribution $P(\dot u)$ is predicted to scale as 
\beq
P(\dot u)\sim \dot u^{-\sf a}\ , 
\eeq
with a very large exponent  ${\sf a}=1$ for the BFM and the ABBM model. On the other hand, the scaling relation  ${\sf a} = 2- \frac2{d+\zeta-z}$ predicts a negative exponent  ${\sf a} = -0.45$ in dimension $d=1$, a quite dramatic deviation from the BFM and MF value.   Remarkably, our simulations confirm this negative value, yielding 
\beq
{\sf a} = -  0.45 \pm 0.05\ .
\eeq

\paragraph{Distribution of spatial extensions.}
We finally consider the spatial extension $\ell$ of an avalanche. Using that $P(\ell)\rmd \ell = P(S) \rmd S$, and $S\sim \ell^{d+\zeta}$, we get
\beq
P(\ell) \sim \ell^{-\sf k}\ , \qquad {\sf k} = d-1+\zeta ~\stackrel{d=1} {-\!\!\!\longrightarrow} ~\zeta=1.25\ .
\eeq
Our numerical data   shown in Fig. \ref{f:PofL2} are in agreement with this scaling relation, yielding 
\beq
{\sf k} =   1.25 \pm 0.05\ .
\eeq
In higher dimensions, the lateral extension of an avalanche is difficult to define, whereas its volume is well-defined. Using scaling arguments equating $P(V)\rmd V = P(S) \rmd S$, $S\sim \ell^{{d+\zeta}}$, and $V\sim \ell^{d}$ we find   
\bea
P(V) &\sim& V^{-{\sf k}_V}\ ,\\
{\sf k}_V &=& 2- \frac{2-\zeta}d\ .
\eea
Explicit values are given in table \ref{tab1}.

\paragraph{Conclusion.}
In this letter, we confronted theoretical results for the distributions of avalanche size, duration, and velocity with numerical simulations.  We confirm the  theoretical results based on scaling arguments, and functional RG calculations to 1-loop order. 
Our comparison goes beyond scaling exponents, validating the full 1-loop scaling functions. 

The model and algorithm  proposed here can be  generalized to arbitrary dimension and long-range interactions. It is computationally more demanding than the standard depinning model due to the presence of multiplicative noise, but it has the advantage that it  allows one to compute more precisely the spatio-temporal extent of an avalanche and to reach the regime of adiabatic driving.
It thus avoids the difficulties and artifacts associated with velocity thresholding. In addition, as its microscopic disorder has the statistics of a random walk   at short scale, it is readily connected to the BFM model.

\acknowledgements
We thank A. Dobrinevski for very useful discussions
and acknowledge support from PSL grant ANR-10-IDEX-0001-02-PSL.
We thank KITP for hospitality and support in part by the NSF 
under Grant No. NSF PHY11-25915.


\ifx\doi\undefined
\providecommand{\doi}[2]{\href{http://dx.doi.org/#1}{#2}}
\else
\renewcommand{\doi}[2]{\href{http://dx.doi.org/#1}{#2}}
\fi
\providecommand{\link}[2]{\href{http://#1}{#2}}
\providecommand{\arxiv}[1]{\href{http://arxiv.org/abs/#1}{#1}}

\end{document}